\documentclass[3p,times]{elsarticle}
\usepackage{preamble/ecrc}
\volume{}
\firstpage{1}
\journalname{}
\runauth{H.Fatima et al}

\jid{procs}

\jnltitlelogo{}

\usepackage{amssymb}
\usepackage[figuresright]{rotating}
\usepackage[acronym]{glossaries}
\usepackage{array}
\usepackage{url} 
\usepackage{hyperref}
\usepackage{breakurl}
\usepackage{float}
\usepackage{xcolor}
\usepackage{multirow}

\newacronym{IoM}{IoM}{Internet-of-Mirrors}
\newacronym{IoT}{IoT}{Internet-of-Things}
\newacronym{IoV}{IoV}{Internet-of-Vehicles}
\newacronym{IoMusT}{IoMusT}{Internet-of-Musical-Things}
\newacronym{5G}{5G}{fifth-generation}
\newacronym{AR}{AR}{Augmented Reality}
\newacronym{CV}{CV}{Computer Vision}
\newacronym{APIs}{APIs}{application programming interfaces}
\newacronym{IoMT}{IoMT}{Internet of Medical Things}
\newacronym{lpwan}{LPWANs}{Low Power Wide Area Networks}
\newacronym{mua}{MUA}{Makeup Artist}
\newacronym{rf}{RF}{Radio Frequency}
\newacronym{ml}{ML}{Machine Learning}

\begin{document}

\begin{frontmatter}


\author{Haneen Fatima\textsuperscript{*}, Muhammad Ali Imran\textsuperscript{*}, Ahmad Taha\textsuperscript{*}, and Lina Mohjazi\textsuperscript{*}}
\address{\textsuperscript{*}James Watt School of Engineering, University of Glasgow, Glasgow, G12 8QQ, UK}

\dochead{}

\title{Internet-of-Mirrors (IoM) for Connected Healthcare and Beauty: 
\newline A Prospective Vision}



\vspace{-7cm}
\begin{abstract}

With the shift towards smart objects and automated services in many industries, the health and beauty industries are also becoming increasingly involved in AI-driven smart systems. There is a rising market demand for personalised services and a need for unified platforms in many sectors, specifically the cosmetics and healthcare industries. Alongside this rising demand, there are two major gaps when considering the integration of autonomous systems within these sectors. Firstly, the existing smart systems in the cosmetics industry are limited to single-purpose products and the employed technologies are not widespread enough to support the growing consumer demand for personalisation. Secondly, despite the rise of smart devices in healthcare, the current state-of-the-art services do not fulfil the accessibility demands and holistic nature of healthcare. To bridge these gaps, we propose integrating autonomous systems with health and beauty services through a unified visual platform coined as the Internet-of-Mirrors (IoM), an interconnected system of smart mirrors with sensing and communication capabilities where the smart mirror functions as an immersive visual dashboard to provide personalised services for health and beauty consultations and routines. We aim to present an overview of current state-of-the-art technologies that will enable the development of the IoM as well as provide a practical vision of this system with innovative scenarios to give a forward-looking vision for assistive technologies. We also discuss the missing capabilities and challenges the development of the IoM would face and outline future research directions that will support the realisation of our proposed framework.
\end{abstract}


\end{frontmatter}

\section{Introduction}  

    \subsection{Background} 
    
    The emergence of the \gls{IoT} has significantly supported the shift towards increased interconnectivity and automation in many sectors such as healthcare, transportation, and smart cities \cite{ma_iqbal_internet_2021}, \cite{al-fuqaha_internet_2015}. The increasing number of ‘things’, internet-connected embedded systems, has opened possibilities for numerous everyday physical objects to interact with each other and achieve shared goals such as enhanced energy efficiency, improved health monitoring, and streamlined supply chains \cite{atzori_internet_2010, miorandi_internet_2012, bariah_prospective_2020}. 

    In recent years, in homes, offices and even industrial contexts, there has been an upsurge of smart environments, appliances and devices that all connect wirelessly over the internet or through local networks. The increased deployment of \gls{IoT} has led to the development of various interconnected frameworks that fall under the umbrella of \gls{IoT} such as the \gls{IoV} \cite{yang_overview_2014} and the \gls{IoMusT} \cite{turchet_internet_2018}. There has been an increased research focus in multiple domains such as transportation and smart cities but there is minimal academic research on the application of \gls{IoT} technologies in the cosmetics and beauty industries. Comparatively, there are ongoing significant research efforts for IoT in healthcare with the rise of telemedicine. However, there is still a need for enhanced and seamless connectivity to expand the reach and accessibility of healthcare systems to the public and enable advanced use-cases. This will not only reduce healthcare staff workload but also improve the patient experience on a holistic level to eventually support better healthcare outcomes \cite{ieeeadmin_integration_2022}. Additionally, this research focus often relates to applications such as emergency management, post-discharge care through health monitoring and patient authentication \cite{haghi_kashani_systematic_2021} but there is a notable lack of academic attention given to specialised and elective healthcare services like dermatology, dentistry, and optometry. Many aspects of elective healthcare services overlap with the cosmetics industry indicating the potential for a unified platform that could encompass both sectors.

    The nature of personalised services in the cosmetics and beauty industry and the elective healthcare sector are visual. An existing smart system that has the potential to support these services is a smart mirror. A smart mirror refers to an intelligent interface that operates as a mirror and can interact with users and display various responses and information. Currently, smart mirror products are becoming increasingly available for assistive beauty and assistive living applications as evidenced by their growing market presence and adoption across sectors such as retail and healthcare \cite{smartmirror_market}. The common features of most smart mirrors include the display of multimedia data and the option for the user to interact with contextual information such as weather conditions and interaction with options such as voice recognition or a touch screen dashboard \cite{alboaneen_internet_2020}. However, despite these features, current smart mirror products are limited to standalone systems, lacking the interconnectivity and comprehensive integration that characterise more mature IoT ecosystems. This gap in connectivity, particularly in the realms of health and beauty, presents an opportunity for innovative solutions that can leverage the visual nature of these services while harnessing the full potential of IoT technologies.

    \subsection{Motivation}

    There is an increasing prevalence and demand for AI-driven services and personalisation in the cosmetics and beauty industry with the integration of technologies like \gls{IoT} and \gls{AR} to provide services such as personalised product recommendations and cosmetic colour customisers \cite{maia_rocklin_IoT_2021}. This demand for personalisation is highlighted in an Accenture survey \cite{noauthor_consumers_2016} which found that 75\% of consumers are more likely to make a purchase when recommendations are tailored to them. Similarly, a Forrester study \cite{noauthor_14_2021} found that 77\% of consumers have chosen, recommended, or paid extra for a brand that offers a personalised service or experience. This trend towards personalisation aligns with the broader healthcare industry's recognition of the critical need for comprehensive, patient-tailored services. As highlighted in \cite{aminizadeh_opportunities_2024}, personalised care that considers patients' unique characteristics, including risk factors and clinical history, can significantly improve treatment efficiency and patient satisfaction. 
    
    Building upon this trend, the beauty and healthcare industries have been quick to adopt new technologies. However, the current landscape of smart beauty and health technologies faces several critical challenges. The current capabilities of commercially available smart beauty products, often termed beauty tech, are limited to information delivery and product customisations with services like portable skin analysers, cosmetic curation apps and skin analysis apps \cite{lee_study_2022}. Although these products provide personalised user experiences, they are not widespread and accessible and serve only a single purpose of service. This indicates the need for a multi-purpose unified platform that can offer personalised services. This fragmentation extends into elective healthcare or aesthetic medicine as well in areas such as dentistry and optometry \cite{baidachna_mirror_2024}. In these areas although there are smart services such as diagnostic support \cite{yang_smart_2020} and mobile app-integrated smart medical instruments \cite{kokol_how_2019}, these also only serve a single purpose and do not provide the clinician or patient with a holistic view of the treatment process.

    Within beauty tech,  smart mirrors have emerged as a promising technology, offering results-driven services such as virtual makeup try-on or product recommendations \cite{rayome_how_2018}. However, drawbacks of beauty tech smart mirrors have been found in consumer interviews conducted in \cite{yum_improving_2022} three main limitations are highlighted, namely; lack of personalisation to individual/consumer needs; lack of research on data exchange between smart devices and the smart mirror (i.e. the \gls{IoT} environment); and limited \gls{AR} features such as real-time virtual try-on. In general, regardless of the smart mirror application, whether it is for the smart home environment or beauty tech, smart mirror products are currently limited to stand-alone systems, failing to integrate with broader IoT ecosystems.

    These limitations mirror the challenges faced in the broader healthcare sector. As highlighted in \cite{aminizadeh_opportunities_2024}, the fragmentation of medical data and the difficulty of generating comprehensive results across different populations hinder the delivery of truly personalised care. Furthermore, while IoT sensors and AI technologies offer the potential for continuous, real-time data collection and analysis, deploying these technologies at scale while ensuring data security and privacy remains a significant challenge. The convergence of challenges in both beauty and healthcare sectors highlights the need for a unified, multi-purpose platform offering advanced personalisation beyond current standalone systems. This platform should integrate data holistically from multiple sources, provide context-aware suggestions, and offer cross-domain insights considering the interplay between health and appearance. It should enable continuous, non-invasive monitoring and seamless data integration while ensuring robust privacy protections. By addressing these aspects, such a platform would offer a more nuanced, adaptive user experience that evolves with individual needs, overcoming the limitations of current fragmented systems in both health and beauty sectors.
    
    \subsection{Contribution}
   To address these challenges and capitalise on the potential of a unified platform for interconnected smart technologies in the health and beauty sectors, we propose a novel concept: the \gls{IoM}. To the best of our knowledge, there is no existing vision or concept of interconnected \gls{IoT}-based smart mirrors. Capitalising on this, this article aims to provide a unified vision for an interconnected system of mirrors coined as the \gls{IoM} under the \gls{IoT} framework. This system will be conceptualised with the existing concept of smart mirrors to support assistive technologies, primarily the health and beauty sector. We will focus on defining the \gls{IoM} ecosystem as well as discussing the state-of-the-art technologies that will enable the existence of the \gls{IoM}. This article seeks to bridge the gaps between an implementation-focused approach to smart mirrors and the need for interconnected services in the assistive health and beauty sectors. Although this paper builds on existing smart mirrors and IoT technologies, it provides a comprehensive and integrated perspective on the inter-operation of all ecosystem components, addressing the gaps in personalised services for the cosmetics and healthcare industries. This perspective is aimed at bridging the gap between the industrial development of smart mirrors and personalised services and the corresponding lack of development in academia, especially in the domains of elective healthcare and beauty tech.

    Overall, the IoM is envisioned to address the unique needs and challenges of the cosmetics and elective healthcare sectors which require the seamless integration of diverse data types to provide personalised and interactive services, which is not typically targeted by IoT frameworks. The IoM differentiates itself by focusing on the interoperability and interconnectivity of smart mirrors within an ecosystem, allowing for enhanced data exchange, real-time analysis, and personalised recommendations. This level of integration is crucial for delivering comprehensive and effective assistive services, ensuring that users receive tailored and holistic care. The user experience directly dictates the methods and approaches that will lead to the realisation of the IoM and this requirement of user-centric criticality also differentiates the IoM from other IoT sub frameworks. Additionally, the IoM can support everyday living, self-care, and non-invasive healthcare by incorporating typically non-smart parts of user lives into the smart city vision, making it distinct from existing smart mirror technologies with limited capabilities.

      \begin{figure*}[h]
            \centering
            \includegraphics{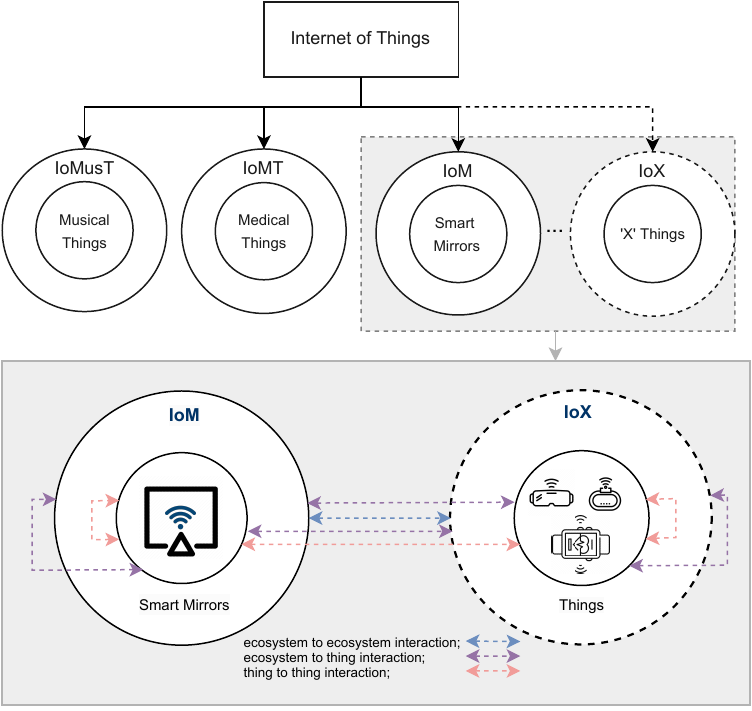}
            \caption{\gls{IoT} example ecosystems and possible interactions.}
            \label{fig:IoTframeworks}
        \end{figure*}
    
    New opportunities may be fostered by the \gls{IoM} for not only assistive beauty but smart living in general, leading to applications and services that can utilise the visualisation capabilities of the \gls{IoM} to connect the physical and digital realms \cite{mohjazi_journey_2023}. However, the realisation and deployment of the \gls{IoM} for end users would also face a number of challenges which will be identified and discussed in this paper. These include low-latency and high-reliability communication protocols for real-time visualisations and services, interoperability and standardisation for the integration of multi-modal content and user privacy and security concerns.
    
    The rest of this paper is structured as follows, Section II defines the concept and ecosystem of the envisioned \gls{IoM}. Section III highlights and surveys existing technologies and works that will support the development and implementation of the \gls{IoM}. Section IV outlines several proposed scenarios related to the \gls{IoM} in order to identify potential capabilities of the system as well as current missing capabilities. Section V describes the main challenges that the development of such a system could face as well as potential solutions and finally, our main conclusions and future research direction are summarised in Section VI.


\vspace{0.3cm}
\section{Concept and Vision}
        \begin{figure*}[h]
            \centering
            \includegraphics{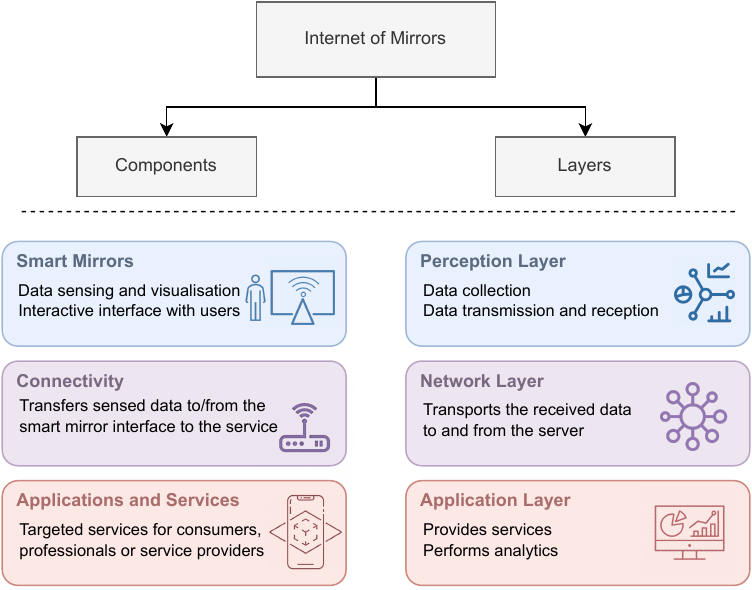}
            \caption{Overall \gls{IoM} ecosystem.}
            \label{fig:IoMlayers}
        \end{figure*}
    \vspace{-0.3cm}
    The \gls{IoM} is a novel concept that can be considered an ecosystem or a subfield of \gls{IoT}. The vision for this ecosystem is for it to interact with other ecosystems and smart systems like the \gls{IoV}, \gls{IoMusT} or \gls{IoMT} \cite{gatouillat_internet_2018} as well as being able to interact with the Things within these systems. In Fig.\ref{fig:IoTframeworks} the possible interactions between \gls{IoT} ecosystems have been visualised where IoX represents an X number of other possible ecosystems within the \gls{IoT} framework. 

     As a system, the \gls{IoM} can be defined as, \textit{“the integration of smart mirror interfaces and related or embedded sensors over a connected network that can support personalised services and user metric visualisation based on sensed data from interactions between the user and smart mirror or interactions between the smart mirrors themselves.”} In this context we consider a smart mirror to be a mirror that can sense, visualise and interact with the user and other components of the \gls{IoM} ecosystem. The ecosystem of these smart mirrors involves collective elements like computing, storage, data exchange and resource management, all of which are garnered by unified protocols and standards.
    
    Considering \gls{IoT} ecosystems in general \cite{boley_digital_2007, hutchison_defining_2012, paolone_holistic_2022}, the \gls{IoM} can have diverse ecosystems depending on the application and its scope. The \gls{IoM} ecosystem, at its fundamental level, involves the users (e.g. consumers, patients, cosmetic and health professionals) and smart mirror interfaces, however from an interconnectivity perspective this system can be categorised into three elements, which are detailed in the subsections below.

    The elements of the \gls{IoM} can also be described in terms of layers similar to a general \gls{IoT} layer structure \cite{ma_iqbal_internet_2021}, where the smart mirrors act as the sensing and transmission layer, and the connectivity of the system can be considered as the network and communication layer. Finally, the services component corresponds to an application layer. The overall \gls{IoM} ecosystem with its components and layers is depicted in Fig.\ref{fig:IoMlayers}.
    
        \subsection{Smart Mirrors} 
        
            \begin{figure*}[ht]
                \centering
                \includegraphics[width = \textwidth]{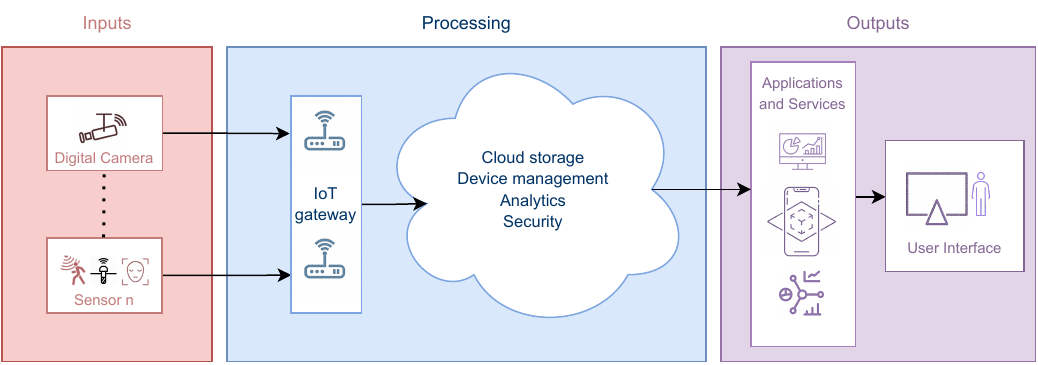}
                \caption{An end-to-end platform of a single smart mirror.}
                \label{fig:smartmirrorprocessing}
            \end{figure*}
            
        The envisioned concept of the \gls{IoM} is for a number of smart mirror devices, along with their sensed data, to be interconnected and have access to other \gls{IoT} devices. The smart mirror devices in the \gls{IoM} system can be considered to inherit the characteristics of Things as its subclass. These characteristics include attributes like sensors, actuators, data collection, analysis and transmission options as well as networking and connectivity features.  
        
        More specifically, a smart mirror will be considered a complete component including all its sensors and processing capabilities, albeit different smart mirrors in the ecosystem can vary from being just a connected interface to a full system with embedded sensing and processing capabilities. However, any smart mirror in the system will have the capability to act as a receiver and/or transmitter by being connected to the relevant local or remote network. A schematic of an end-to-end platform of a single smart mirror is shown in Fig. \ref{fig:smartmirrorprocessing}.
        
        With the increasing interest in existing smart mirrors \cite{kasparian_your_2021}, the integration of the envisioned IoM could enable many new opportunities not only for the cosmetics and health industries but also for other domains such as smart homes and smart education. The visualisation aspect of the smart mirror interface also has the potential to transform user accessibility for consumers and patients by offering a more enhanced and innovative health and beauty experience. 
       
        \subsection{Connectivity}

        \begin{table*}[h]
        \caption{Examples of 5G and LPWAN integration in smart mirrors.}
        \small
        \begin{tabular}{|m{2.7cm}|m{3.3cm}|m{9.2cm}|}
                \hline
                Wireless Technology & Feature & Explanation\\
                \hline
                5G & Cloud Connectivity & Connection with \gls{5G} optimised cloud services would allow smart mirrors to access and store data efficiently and quickly.\\ \cline{2-3}
                & App Integration & Built-in \gls{5G} compatible apps can utilise \gls{5G}'s high speeds and low latency communication for the delivery of high quality services and analytics.\\ \cline{2-3}
                & WiFi Hotspot & The smart mirrors can act as wireless access points which would allow other devices to connect to \gls{5G} networks through them.\\ \cline{1-3}
                LPWANs & Asset Tracking & Tracking of the location of physical assets and other personal items would allow users to locate their items quickly and easily.\\ \cline{2-3}
                & Energy Management & Integration with LPWAN-enabled energy management systems would enable smart mirrors to monitor and adjust energy consumption in real time.\\ \cline{2-3}
                & Smart Home Integration & Integration with smart home systems that use LPWANs would enable smart mirrors to communicate with other devices in the home, such as smart locks, smart thermostats, and security systems.\\ \cline{2-3}
                & Remote Monitoring & Users could monitor their mirror's usage and performance remotely to allow for ease in maintenance and troubleshooting.\\ \cline{2-3}
                \hline
        \end{tabular}
        \label{tab:5GandLPWANinsmartmirrors}
        \end{table*}
        
        The connectivity of the \gls{IoM} ecosystem is fundamental to its operation and development. To support the envisioned multi-directional communication between smart mirrors and from smart mirror to the user (and vice versa), appropriate network protocols and standards must be considered. Depending on the application and domain, different local and remote wireless communication protocols can be deployed. The connectivity of the \gls{IoM} would facilitate the processing of computationally heavy commands and requests to off-load the computation from the embedded micro-controller in the smart mirror. Although the applications of the \gls{IoM} in the cosmetics industry are not time critical in comparison to health applications, real-time use cases such as remote consultancy would still require the network to fulfil low latency, high reliability and synchronisation requirements. The connectivity element of the IoM will not only enable interconnection between IoM smart mirrors and devices but could also facilitate additional services to support the user experience. Table \ref{tab:5GandLPWANinsmartmirrors} provides an overview of two key wireless technologies, 5G and \gls{lpwan}, and possible features that could be enabled by them in the \gls{IoM}.

        \subsection{Applications and Services}
\begin{figure*}[ht]
                \centering
                \includegraphics{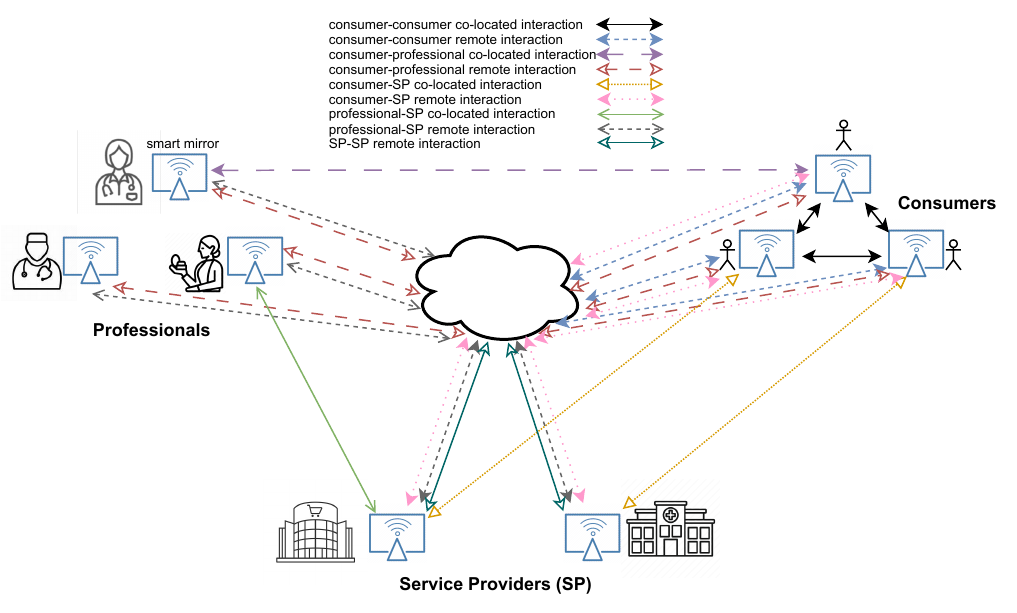}
                \caption{Interaction possibilities between consumers, professionals and service providers.}
                \label{fig:IoMinteractions}
        \end{figure*}

        Whether the application domain for the \gls{IoM} is assistive health, beauty or living, there are several services that can be developed from the sensed and communicated data from the smart mirrors and their connectivity features. These services can target various stakeholders including consumers (beauty and cosmetic customers, patients), professionals (cosmeticians, dermatologists, doctors) and service providers (brands, stores, clinics). To establish an efficient interactive interface of the smart mirrors, these services will need to consider real-time computation and requirements. Fig.\ref{fig:IoMinteractions} shows the possible remote and co-located interactions between different users in the \gls{IoM} ecosystem.

\subsection{Novel Features}
    
   The IoM is envisioned as a unified platform that builds upon and complements existing technologies, designed to work seamlessly with current smart systems while offering enhanced capabilities through interconnectivity. As a framework, the IoM concept presents several novel features that establish its unique value proposition:
    
    \begin{enumerate}
            \item \textbf{Unified Health and Beauty Platform:} Unlike existing solutions that typically focus on single-purpose or single-domain services, IoM provides a comprehensive platform that recognises and leverages the interconnection between these domains. This integrated approach enables more holistic and personalised recommendations that consider both health and aesthetic factors.
        
            \item\textbf{Specialised Interconnected Ecosystem:} Unlike general IoT systems, the IoM creates a specialised network of smart mirrors designed specifically for health and beauty applications. This mirror-centric approach allows for unique capabilities such as real-time visual analysis, augmented reality overlays for beauty treatments, and a consistent user interface across different locations (home, clinics, retail spaces). The interconnected nature of these mirrors enables a seamless experience as users move between different environments, with their personalised data and preferences following them.
            
            \item \textbf{Continuous, Non-invasive Monitoring:} By leveraging advanced computer vision and AI technologies, the IoM framework complements existing wearable sensors by offering additional non-invasive monitoring capabilities. This approach can augment data from wearables, potentially increasing overall user compliance and providing a more comprehensive dataset for analysis. The combination of wearable data and IoM-derived information offers a richer, more nuanced picture of user health and beauty metrics.
            
            \item \textbf{Context-aware Personalisation:} The IoM system integrates visual data with environmental factors and historical user information to provide highly personalised, context-aware recommendations. This multi-faceted approach to data analysis enables more accurate and relevant suggestions than current standalone systems.
        
            \item  \textbf{Seamless Professional Integration:} IoM facilitates easier integration of professional expertise into daily routines. It can serve as a platform for remote consultations with healthcare providers or beauty professionals, bridging the gap between at-home care and professional services.
            
            \item \textbf{Adaptive Learning:} Through its interconnected nature, the IoM system can learn from aggregated, anonymised data across its network. This enables the system to continuously improve its recommendations and adapt to changing trends and user needs.
    \end{enumerate}
    
    These novel features address the limitations of current fragmented solutions in health and beauty technologies, as discussed in Section 1.2. By providing a unified, interconnected platform, IoM aims to offer a more comprehensive, personalised, and effective approach to health and beauty care. While these features present significant potential for advancing health and beauty care, it is important to acknowledge that realising the full potential of the IoM framework involves addressing several significant research challenges. The seamless interactions and data exchange across the IoM ecosystem necessitate innovations in various technological domains. These include advanced communication and network technologies, robust security and privacy measures, enhanced interoperability protocols, standardisation efforts, and scalable AI and machine learning algorithms. Addressing these challenges forms a core motivation for the IoM vision, driving forward research and innovation across multiple disciplines. The pursuit of solutions to these open questions may contribute not only to the specific domain of health and beauty care but also to the broader fields of IoT, AI, and connected technologies. Section 5 of this paper provides a more detailed examination of these challenges and potential approaches to address them, underscoring the interdisciplinary nature of the research required to fully realise the IoM framework.

\section{Enabling Technologies}

In this section, a list of state-of-the-art technologies related to the \gls{IoM} is presented. We aim to describe the results of various application domains that lead to the conceptualisation of the \gls{IoM}. A top-level overview of the presented technologies and their potential integration in the \gls{IoM} is shown at the end of this section in Table \ref{tab:IoMtechintegration}.

    \subsection{Standalone Smart Mirrors}
        The current usage of the term "smart mirror" refers to a mirror that is enhanced with various features, such as touchscreens, voice assistants and other digital displays \cite{alboaneen_internet_2020}. Smart mirrors are designed to offer users a more interactive and personalised experience while using the mirror. In the late 2000s and early 2010s, research related to smart mirrors referred to live webcam footage and implemented services on the live feed as a smart mirror \cite{iwabuchi_smart_2009}, \cite{rahman_augmented_2010}. 
        
        One of the earliest examples of a modern-day smart mirror is the "Mirror, Mirror on the Wall" project by Michael Teeuw, which was first introduced in 2014 \cite{noauthor_michaelteeuwnl_nodate}. Teeuw used a Raspberry Pi computer, a monitor, a two-way mirror, and various other components to create a customised mirror with a digital display that could show news, weather, and other information. Similar to this project, the majority of literature on smart mirror development focuses on the mirror’s technical implementation and hardware of smart mirrors with embedded microcontrollers namely Raspberry pi \cite{sahana_raspbian_2023} and occasionally Arduino Uno \cite{latif_real-time_2019}.  The most common features implemented even in more recent studies are live weather, time, calendar and music display integrated either through a web API or through a connection to a smartphone \cite{prasad_making_2021, barbadekar_design_2023, manjrekar_smart_2022, subramanya_chari_raspberry_2023}. 
        
        In recent years, there have been some increased research efforts within this domain to develop and integrate new features such as the use of Infrared (IR) sensors for user presence detection \cite{jin_design_2018}, gesture control and tracking for music control \cite{prajapati_interactive_2022} as well as WiFi connectivity \cite{sun_design_2018} for mobile synchronisation with the digitally displayed data \cite{athira_smart_2016}. There have also been efforts in the development of plugins and web APIs to support smart mirror applications \cite{gold_smartreflect_2016} \cite{akshaya_smart_2018}. There has also been an increase in the number of commercially available smart mirrors for both assistive beauty and assistive living applications. There are many commercial options of stand-alone mirrors for assistive beauty purposes as well as assistive health. Commonly available smart mirrors with their main features have been shown in Table \ref{tab:commercialSMs}.    

        \begin{table*}[h]
         \caption{Commercially available smart mirrors and their features.}
         \centering
         \normalsize
            \begin{tabular}{|m{6cm}|m{8cm}|}
                \hline
                \textbf{Name} & \textbf{Features}\\
                \hline
                \hline
                Sephora Mirror \cite{bargh_b_sephora_2019} & QR code generation for in-store and online makeup, skincare and fragrance recommendations\\ [2ex] \hline
                HiMirror \cite{1} & Adjustable lighting, skin analysis, product recommendation, cosmetics tutorial\\ [2ex] \hline
                MemoMi \cite{2} & Image capturing and storage,\gls{AR} eyewear and clothes, Smartphone and SNS sharing\\ [2ex] \hline
                Charlotte Tilbury MagicMirror \cite{3} & Augmented makeup, importing images\\ [2ex] \hline
                Shiseido Digital Counseling Mirror \cite{4} & Product information, application instructions, skin check, smartphone compatibility\\ [2ex] \hline
                Panasonic New Magic Mirror \cite{5} & Skin condition identification, virtual try-on\\ [2ex] \hline
                Modiface \cite{6} & Product recommendation, virtual try-on\\ [2ex] \hline
                Lumini Kiosk V2 \cite{7} & Skin analysis, product recommendation, skin comparison with other users\\ [2ex] \hline
                Icon AI Z Mirror \cite{8} & Adjustable lighting, skin analysis, product recommendation, virtual try-on\\ [2ex] \hline
                Morror \cite{9} & Adjustable lighting, cosmetics tutorial, weather information\\ [2ex] \hline
                Simplehuman Sensor Mirror Pro \cite{10} & Smart lighting\\ [2ex] \hline
                Kohler Verdera Voice Lighted Mirror \cite{11} & Voice activation integrated with Alexa and Google assistant\\ [2ex] \hline
                iHome Reflect Smart Vanity Mirror \cite{12} & Lighting and speaker control with smartphone\\ [2ex] \hline
                Capstone Connected Home Smart \cite{13} & Integration with other smart home devices, built-in speaker, voice activation\\ 
                [2ex] \hline
            \end{tabular}
            \label{tab:commercialSMs}
        \end{table*}

        The commonality between all these aforementioned commercially available systems is the lack of features in a standalone smart mirror system. However, the smart mirrors catered for cosmetics and beauty have relatively more varied features than those targeted for assisted living use. Comparatively, in academia, there are limited features and services that cater to assistive beauty because the main domain and application of smart mirrors are for assisted living \cite{ghazal_mobile-programmable_2017} \cite{nadaf_smart_2021}. A relevant example is the development of a chatbot integrated into the smart mirror for the elderly \cite{prajugjit_friendly_2022}. Similarly, in Yang \textit{et al} \cite{yang_error-resistant_2022}, an error-resistant movement detection algorithm is integrated into the smart mirror for increased accuracy for pose estimation for the elderly.  Pose estimation from smart mirrors has also been considered in a fitness context where Hippocrate \textit{et al} have developed a smart mirror to automatically detect and count repetitions of exercises \cite{hippocrate_smart_2017}. This opens up possibilities for the accurate detection of joint and/or mobility-related health conditions for the elderly as well as guided fitness coaching \cite{gomez-carmona_smiwork_2017}. Specifically for assistive health, the main focus and use of smart mirrors are currently services like health monitoring \cite{muneer_smart_2020} with the processing of multi-modal biomedical data for hospital or clinic settings \cite{mIoTto_reflecting_2018}. 
        
        Smart mirror applications and services also include user-specific recognition systems for personalised user metrics and personalised recommendations. This includes facial recognition where facial expressions have been analysed and classified to detect premenstrual syndrome symptoms \cite{ahn_smiley_2022}, deliver personalised content \cite{kuruppu_smart_2021}, control lighting based on mood \cite{yang_intelligent_2018} and security for smart living applications by verification of user identity through facial and voice authentication \cite{njaka_voice_2018} \cite{uddin_mirrorme_2021}. These recognition and detection methods have the potential to increase personalisation in the cosmetics industry and give accurate product recommendations such as in Nguyen \textit{et al} where a makeup recommendation and synthesis method is proposed \cite{nguyen_smart_2017}. Although the synthesis of the product is over a still image, \gls{AR} technologies can be combined with this method to enable personalised and accurate virtual try-on in real-time. An example for skincare applications exists in \cite{dragomir_smart_2020} where a smart mirror is used for skin-type classification and identification, although the system only gives the user the result of the identification there is potential for product recommendations to be integrated in the beauty tech context as well as healthcare context for skincare product recommendation.
        
        Since the developed prototypes in most mentioned works have two or three limited features, the processing is feasible with a server integrated on a stand-alone smart mirror. However, future larger-scale smart mirrors or \gls{IoM} ecosystems comprising of inter-connected smart mirrors with several integrated features, will entail the need to deploy distributed computing approaches. The inclusion of increased features also requires energy and power consumption considerations, this has been preliminarily considered by Rat \textit{et al} \cite{rat_energy_2022} who presented a comparison of the power consumption of different operating systems and different smart mirror modules, such as voice recognition and live weather updates, by enabling and disabling these features. Solutions for mitigating the effects of power-intensive tasks are presented as a trade-off between processing on the cloud and processing latency.
        
        Additional research gaps that are apparent from the state of the art of current smart mirrors include low-quality resolution of the smart mirror due to the nature of its hardware construction. Current studies construct the smart mirror by the addition of a reflective surface such as an acrylic sheet to an LED display so as to act as a two-way mirror. However, the optical resolution of the reflective surface is comparatively lower than a regular glass mirror, hence there is potential to optimise and improve the fundamental composition of the smart mirror hardware components such as incorporating anti-reflective coating, using a glass 2-way mirror \cite{wouk_kris_samsung_2015} or the use of a transparent OLED display. Finally, there is a qualitative research gap in the smart mirror research area due to the lack of empirical evidence on the effectiveness of smart mirrors in improving the user experience through its smart living applications \cite{yum_improving_2022}.

\begin{figure*}[h]
            \centering
            \includegraphics[width=0.7\textwidth]{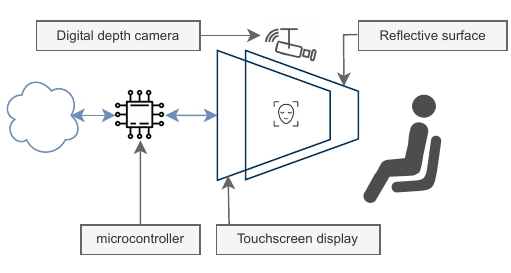}
            \caption{A simple \gls{IoT}-enabled smart mirror schematic.}
            \label{fig: single smart mirror}
        \end{figure*}

        Capitalising on the aforementioned studies, we illustrate a typical modern-day stand-alone smart mirror in Fig.\ref{fig: single smart mirror}.

        \subsection{Skin Imaging Methodologies}
        
        Skin imaging technologies have advanced significantly in recent years. Some common advanced image processing techniques that are being used in the analysis of skin imagery include classification, feature extraction and segmentation. Skin imaging technologies can serve as a powerful tool for disease detection in assistive health scenarios \cite{owda_early_2022}. By identifying and classifying skin features, these technologies can assist in the early detection and diagnosis of various skin conditions, including melanoma, eczema, and acne. Early detection is crucial in managing these conditions and can significantly improve treatment outcomes. For instance, melanoma detected and treated early is almost always curable, while if left untreated, it can quickly spread to other parts of the body and become fatal \cite{owda_early_2022}.

        \begin{itemize}
            \item \textbf{Classification and feature extraction} are integral components of image processing that work in collaboration to analyse skin images. First, specific skin characteristics, such as texture, colour, and shape, are identified and quantified using feature extraction techniques.  These features are then categorised based on predefined criteria through classification. Both processes are commonly facilitated by \gls{ml} algorithms, with convolutional neural networks (CNNs) being a popular choice due to their effectiveness in image analysis \cite{sharma_analysis_2018}. An example of this is in \cite{alarifi_facial_2017} where the authors use CNNs to classify non-clinical skin imagery, specifically facial skin patches, by normal skin, spots and wrinkles. The study found potential for increased accuracy if a larger dataset was collected. Similarly, in \cite{bianconi_personal_2017} skin imagery of the forearm is used to extract features of skin texture and use classification for personal identification. However, in this study clinical imagery was collected from a dermoscope and a capacitive imaging device. By comparing both studies, both non-clinical and clinical imagery like capacitive imaging have the potential to be integrated into a larger multi-modal dataset to create a model with increased accuracy. 
        
             \item \textbf{Capacitive imaging }is often used for skin hydration measurements, this can be combined with classification techniques such as in \cite{zhang_skin_2020} where the classifications of different hydration levels were used to identify different regions of the body. This non-invasive method to monitor skin hydration levels can contribute further insights into the skin's overall health analysis and can potentially accurately guide skincare treatments recommended by the envisioned \gls{IoM}.
        
             \item \textbf{Segmentation} can also be used as a pre-processing stage before feature extraction to separate the skin image into different regions or objects to isolate relevant areas. This can specifically target areas like lesions for diagnosis support and treatment planning. This can be seen in \cite{jain_computer_2015} where the authors segment the input image to isolate skin lesions and extract different features like asymmetry, colour and size for melanoma detection. The combination of segmentation, feature extraction and classification techniques for skin image processing allows for more accurate detection of skin conditions and opens opportunities for increased accuracy not only in diagnosis but also in skincare treatment recommendations.
        
             \item \textbf{Disease detection} has also benefited from the aforementioned image-processing techniques, such as in \cite{alkolifi_alenezi_method_2019} and \cite{bhadoria_model_2020}, where the authors identify different skin conditions such as melanoma using pre-trained CNNs. In a more recent study \cite{sharma_skin_2022}, a similar methodology is built upon to incorporate the detection of other skin conditions like eczema, psoriasis and acne with a larger dataset. There are also studies related to the detection of acne such as in \cite{shen_automatic_2018} where the acne is categorised by a seven-classifier. Likewise in \cite{zhao_computer_2019}, feature extraction and classification are used for acne detection by its severity but the study also integrates this service into a mobile application which can detect acne from a self-taken image. Moreover, multispectral imaging techniques \cite{kuzmina_towards_2011} are also particularly effective in disease detection \cite{prigent_multi-spectral_2010} \cite{kuzmina_multi-spectral_2011}. These techniques enable an analysis of skin properties below the surface that are not visible to the naked eye, providing a deeper understanding of various skin conditions \cite{spigulis_multispectral_2010}.
        
             \item \textbf{Wearable skin sensors and dermatoscopes} have also made skin imaging more accessible and practical \cite{ahmad_tarar_wearable_2020}. These devices can capture high-resolution images of the skin, allowing for detailed analysis and assessment \cite{huang_recent_2022}. This is particularly useful in remote healthcare scenarios, where patients can use these devices to capture images of their skin \cite{huang_new}, which could then be analysed by remotely connected dermatologists or AI-enabled detection systems embedded into a prospective \gls{IoM} for diagnosis and treatment planning.
             
        \end{itemize}
        
        Overall, the integration of skin imaging technologies with smart mirrors and the foreseen \gls{IoM} could provide users with a more comprehensive view of their skin's condition and enable them to make informed decisions about skincare and makeup products. These technologies are expected to assist in the diagnosis process by offering detailed and labelled images and analyses of the skin, along with suggested diagnoses. For example, by utilising classification and feature extraction techniques, dermatologists will be able to identify and quantify specific skin characteristics from the image, such as texture, colour, and shape. This is anticipated to aid in the diagnosis of various skin conditions and guide treatment decisions. Furthermore, the use of segmentation techniques will allow dermatologists to focus directly on isolated relevant areas in skin images, such as skin lesions, automatically highlighted by the future IoM. AI-driven skin analysis is projected to significantly reduce diagnosis time and improve the accuracy of diagnoses. This efficiency will be particularly beneficial in dermatology, where timely and accurate diagnoses are critical. Future trends in skin imaging may include more accurate diagnosis, early detection of skin conditions, and personalised treatment planning. However, these advancements also bring forth challenges related to image quality, data privacy, and regulatory compliance, which would need to be addressed to ensure the successful implementation of these technologies.  
        
    \subsection{Real-Time Health Monitoring}
        
        Real-time health monitoring is a major area of focus within assistive health. By combining the capabilities of non-invasive sensing, data analytics and instant feedback, the \gls{IoM} can be integrated with real-time health monitoring to improve the patient experience in a convenient and non-intrusive way as well as ease the consultancy process for health professionals in terms of early detection, prevention, and better management of health conditions.

        Within real-time health monitoring, vitals tracking can provide a wealth of data for diagnostics and health assessments. We foresee that Non-contact sensing technologies like  \gls{rf} sensing, photoplethysmography (PPG) and infrared thermography can be integrated in the IoM to continuously monitor critical physiological parameters to support assistive health applications.

        \begin{itemize}
        
         \item \textbf{PPG}, for instance, uses light to measure changes in blood volume, enabling the tracking of heart rate, respiratory rate, and oxygen saturation levels \cite{sinhal_overview_2020}. The technique is particularly advantageous for continuous, non-invasive monitoring and can be implemented using simple, cost-effective hardware. These metrics can also be tracked using a digital camera such as in \cite{villarroel_non-contact_2017} where the authors use the colour information of patient video footage to detect the blood pulse volume to estimate these metrics. In \cite{ghanadian_machine_2018} this method of heart rate variability tracking using an RGB camera is improved by separating the source from the noise in the video recording of the subject. Similarly to cameras, microphones, which are often integrated into many smart mirrors, can also be used to capture data on breathing patterns which can be used to estimate heart rate and other indicators of health.

         \item \textbf{Infrared thermography} is also a non-contact imaging technique that provides a different approach by capturing the infrared radiation emitted by an object using an IR camera, typically in the form of heat. The infrared radiation is then converted into an electronic signal, which is processed to produce a thermal image on a video monitor and perform temperature calculations \cite{jagadev_human_2020}. Since the skin's temperature distribution can be influenced by a variety of physiological and pathological factors, thermal images can provide valuable diagnostic information such as patterns of inflammation, vascular activity and metabolic processes \cite{verstockt_skin_2022}. This information can contribute to the early detection and diagnosis of various skin conditions like psoriasis, dermatitis and localised scleroderma \cite{vergilio_evaluation_2022}. Monitoring changes in skin temperature over time with infrared thermography can also provide insights into the progression of diseases and the effectiveness of treatments. A recent study uses this sensing technology to monitor the respiration rate (RR) by using \gls{ml} to differentiate normal and abnormal breaths by predefined thresholds and obtain the breaths per minute, without any manual intervention as compared to traditional RR monitoring methods \cite{jagadev_non-contact_2020}. This study shows the potential for non-contact sensing to monitor and predict not only skin conditions but also overall health conditions since respiration is a vital physiological process and can be an indicator of different breathing disorders. 

         \item \textbf{\gls{rf} sensing} encompasses a broad category of technologies that use radio waves to detect changes or events. Shah and Fioranelli provided a comprehensive review of \gls{rf} sensing technologies, highlighting their potential in assisted daily living and healthcare settings as a less intrusive means of health monitoring \cite{shah_rf_2019}. Radar sensing, as part of the broader category of \gls{rf} sensing technologies, also offers significant potential for non-contact vitals tracking. We anticipate its integration into the smart mirror sensing system in the \gls{IoM} ecosystem with a device such as the Xethru X4 Impulse-Radio Ultra-Wideband (IR-UWB) radar system on a chip \cite{thullier_systematic_2022} which can be used to monitor body movements, including breathing and heart rate, without the need for physical contact or sensors \cite{kebe_human_2020}. This can be achieved by measuring low-power radio waves emitted towards the user and measuring the frequency shift of the waves as they bounce back from the user's body. Frequency changes in the waves can be measured and used to detect body movements, including chest movements associated with breathing and heart rate. \gls{rf} sensing can also be used to monitor more complex physiological parameters by detecting small displacements of the skin surface associated with arterial pulsation, providing an indirect measure of heart rate and blood pressure. This could be particularly useful for the continuous monitoring of patients with cardiovascular disease \cite{gao_new_2017}. Further advancements in \gls{rf} sensing are seen in the development of high-resolution chipless tags for \gls{rf} sensors, as presented by Abbasi \textit{et al} in \cite{abbasi_high-resolution_2020}. These tags enable more detailed and accurate data collection, enhancing the potential for detailed health monitoring and diagnostics. Comparatively, a more unique application of compact \gls{rf} sensing is explored by Zheng \textit{et al} in their concept of "V2iFi" vehicle vital sign monitoring \cite{zheng_v2ifi_2020}. This concept exhibits the flexibility and adaptability of \gls{rf} sensing methods \cite{cui_interacting_2023}, suggesting their applicability in diverse settings, including smart mirrors and the anticipated \gls{IoM}. 

        \end{itemize}

        In our envisioned \gls{IoM}, aside from integrating real-time health monitoring through cameras and sensors, the smart mirror will also have the capability to utilise data from wearables to provide a holistic view of an individual user's health metrics \cite{loncar-turukalo_literature_2019}. Future wearable devices equipped with \gls{rf} sensors are expected to range in applications from monitoring respiratory patterns \cite{sharma_skin_2022} to detecting tremors in patients with Parkinson's disease \cite{abirami_b_-body_2021}. Parallel to wearables, the emerging paradigm of 'skintronics' or skin-interfaced electronics, which will merge with the user's skin, holds high potential for achieving increased accuracy in health metrics monitoring \cite{zhu_skin-interfaced_2023}. Additionally, personalised health monitoring is envisioned to be enhanced by smart textiles \cite{weng_smart_2016}, where circuits and systems will be integrated into everyday clothing \cite{lin_fiber-based_2015}. Haptic feedback technologies are also expected to embed tactile experiences directly into these textiles, offering users an enhanced, multi-modal interaction \cite{huang_application_2020}. These technologies are not just for monitoring but also for interactive alerts and warnings. We anticipate that these wearable technologies will interact seamlessly with the future \gls{IoM} to offer real-time feedback, tailored health recommendations, and tools for proactive health management.
        
    \subsection{Computer Vision}
        Computer vision is a significant area of focus for makeup-related applications \cite{treepong_development_2018} and is often used as a final stage after user feature detection and implementation of a recommendation system. In our envisioned \gls{IoM}, computer vision will have the capability to augment cosmetics or provide \gls{AR} experiences on user images. These future systems, relying on live camera footage and infrared cameras, will be able to superimpose different makeup products virtually on the user in real-time, offering a highly interactive and engaging user experience.
        
        Several studies exist on the augmentation of cosmetics or \gls{AR} on user images with the earliest in 2009 \cite{iwabuchi_smart_2009} \cite{rahman_augmented_2010}, where the authors propose a smart mirror system composed of a live web camera footage integrated with an IR camera for position tracking, to augment different makeup products on user virtually. Similarly, in recent years more sophisticated and user-centric approaches are emerging such as in \cite{soares_borges_virtual_2019} where makeup is augmented through a tactile experience by allowing the user to use their finger to simulate a physical applicator. Their initial prototype could accurately detect touch from 2.2mm and also achieved real-time interactive performance. In \cite{kim_color_2020} however, the authors identify the limitation of colour reproduction in virtual make-up since previous studies focused on the recommendation of the makeup style so they propose an accurate lipstick colour reproduction method based on CNNs. 
        
        For further improvements in virtual makeup systems, in \cite{perera_virtual_2021} a method to increase personalisation is proposed by considering a more holistic recommendation system that goes beyond facial features.  By factoring in classifiers like skin tone, attire, colour, and hair features, this system offers hyper-personalised makeup recommendations, attuned to individual preferences and unique physical attributes. In \cite{cho_makeup_2019}, a similar system is designed but a geometric approach is taken instead based on colour and shape parameterisation. Similarly, in \cite{treepong_makeup_2018} \gls{AR} technology has also been utilised in enhancing makeup creativity  where makeup tools that provide tangible interaction and generate colour on the user have been developed. It also synchronises with the user's movement in real time. The study also conducted a qualitative evaluation to evaluate the users’ satisfaction, which indicated that by using this system, users can find a suitable makeup style for themselves and get help increasing their makeup creativity.

        The realm of \gls{AR} in our proposed \gls{IoM} ecosystem is not just limited to virtual makeup systems but can also entail the integration of computer vision in retail scenarios, as seen in  \cite{aly_toward_2021} which demonstrates the potential of these technologies in broader applications, such as automatic body measurements for apparel selection. The versatility of \gls{AR} and computer vision has been showcased in diverse applications that are relevant to the \gls{IoM}, from enhancing fitness routines through \gls{AR} technology \cite{ueta_improving_2022} to the accurate detection and synthesis of makeup features \cite{ren_new_2019} as well as 3-D tracking of shoes with smart mirrors \cite{eisert_3-d_2008}. 
        
    \subsection{Personalised Recommendation Systems}

        For assistive health and beauty services, recommendation systems are essential to providing product, treatment and routine suggestions to the user. The first step towards personalised recommendations in the \gls{IoM} is the detection and analysis of user-specific features, this can include skin parameters, facial feature details, vitals detection or even makeup detection such as in \cite{alzahrani_deep_2021}, where deep learning models are used to detect makeup. These models are trained to recognize subtle changes in the skin, such as the application of makeup, and can therefore provide a detailed analysis of the user's current makeup status. This information forms the basis for personalised makeup recommendations, highlighting products that complement the user's current look or suggesting changes to improve their appearance. These services are offered by many large cosmetic brands such as a foundation shade finder and recommender by L'oreal Paris \cite{loreal}, Maybelline \cite{maybelline} and Nars \cite{nars}.
        
        The results from a detection algorithm can consequently be used to present a corresponding recommendation like the authors in \cite{wang_face_2016} who detect and classify eye shapes to give augmented real-time eyeliner style guidance and recommendations. Most recommendation systems for makeup applications are based on facial imagery datasets \cite{hu_framework_2016} and a corresponding product database. There is a lack of data from professionals being incorporated into the recommendation models. This is highlighted in \cite{lee_cognitive_2022}, the authors propose a cognitive knowledge-based system for hair and makeup recommendations that combines user feature classification with a database of professional knowledge. The system uses \gls{ml} algorithms to identify and categorise facial features, such as eye shape and skin tone. It then matches these features with professional recommendations to provide personalised makeup and hairstyle suggestions. The inclusion of professional expertise in this process significantly enhances the quality of the recommendations, providing users with advice that aligns with current beauty standards and trends. 
        
        Further developments to existing makeup recommendation systems were made in\cite{gulati_beautifai_2022}, which proposes a system that offers occasion-oriented makeup recommendations. By considering the user's current occasion, the system provides makeup suggestions that are appropriate for the situation. This ensures that the recommendations are not only personalised to the user's features but are also relevant to their immediate context. The novelty in the work includes multi-region-wise makeup recommendations, incorporation of occasion context, real-time makeup previews and continuous makeup feedback. Additionally, 3D face recognition \cite{li_comprehensive_2022} can also be utilised in recommendation systems to provide a detailed and accurate representation of the user's face \cite{liu_facial_2015}. This detailed facial map can be used to provide more precise recommendations, taking into account the unique contours and features of each user's face.
        
        We envision that the integration of recommendation system in the IoM will include the incorporation of user feedback with deep learning methods \cite{dau_recommendation_2020}. These systems can learn from the user's past behaviour and preferences, providing recommendations that align with their user-specific needs. For example, if a user frequently chooses a specific makeup look, the system will prioritise similar recommendations in the future. In addition, by incorporating user feedback, these systems can continually refine and adjust the recommendations based on the user's satisfaction with previous suggestions. This type of corrective learning creates a more responsive and adaptive recommendation system, leading to a highly personalised and enhanced user experience. The use of corrective learning approaches can also help address potential biases in the recommendation system. For instance, biases may occur when certain products are recommended more often due to higher ratings or popularity. By incorporating user feedback, the system can learn to correct these biases, ensuring that recommendations are truly personalised and not influenced by external factors. Socially aware recommendation systems are another significant development in the realm of data analytics \cite{villegas_characterizing_2018}. Such systems consider the social context of the user, taking into account their social network data to provide recommendations. For instance, a socially aware recommendation system can consider the user's friends' product preferences or their reactions to certain beauty products. This social context can significantly enhance the recommendation quality, providing suggestions that are more likely to resonate with the user's social environment.
            
       \subsection{Wireless Connectivity in \gls{IoT}}
     
        \gls{IoT} is an integrated system of cutting-edge technology that enables connections between people, devices, platforms and solutions over the Internet \cite{ma_iqbal_internet_2021}. There are many connectivity aspects that an \gls{IoT} system must consider. Specifically for an \gls{IoM} connected to other \gls{IoT} devices, it is essential for the connectivity of the system to be in accordance with its computational load and latency requirements. 
        
        The specific communication enablers that would be relevant to \gls{IoT} applications in assistive health and beauty are dependent on the specific application and range of the network \cite{chettri_comprehensive_2020}. This can be seen in the current wireless communication protocols deployed for different smart environments \cite{shafique_internet_2020}, such as smart homes which are often connected using the WLAN WiFi due to the short-range nature of the application. Smart health, on the other hand, also uses 3G, 4G and satellite communications since the applications can be anywhere from a local hospital setting to a global remote healthcare setting \cite{al-hawawreh_threat_2021} \cite{jamoos_low_2023}. Moreover, \gls{lpwan} could be attractive for specific \gls{IoT} applications, particularly those requiring wide coverage and low power consumption, like environmental monitoring or asset tracking \cite{khanh_wireless_2022}. Meanwhile, wireless sensor networks (WSNs) and local area networks (LANs) could benefit more localised applications, offering high data rates over a smaller area \cite{moraes_wireless_2017}.

        Furthermore, \cite{ding_IoT_2020} have classified \gls{IoT} applications based on end-user type, either machine-oriented or human-oriented, and based on data rates. Applications similar to our proposed system include health risk detection systems which lie in the middle of both end-user type and is on the higher end of the data rate spectrum with data rates between tens of Mbps to tens of Gbps. Intelligent shopping applications, which hold similarities to the requirements of the data analytics in our system, require lower data rates within the range of 1Mbps to 10Mbps. Despite the lack of academic outlook and industrial implementation of wireless connectivity in the cosmetics industry, the analogous retail industry has existing examples of wireless connectivity being incorporated into their application. A recent example is the collaboration between Huawei and the  Square shopping centre in Camberley to become the first fully \gls{5G}-enabled retail destination in the UK \cite{ismailjee_uk_2020}. The centre covers a 460,000 sq ft area and attracts over 8 million visitors a year, indicating that the \gls{5G} network employed has the potential to support a system with similar requirements and demands, such as the cosmetics industry and beauty tech applications. 

        For larger-scale systems or systems that require large bandwidth and low latency, \gls{5G} wireless communication, especially its millimeter-wave (mmWave) technology, can address the limitations of previous generations. The mmWave spectrum offers vast amounts of bandwidth, facilitating high data rates and acting as a key component of \gls{5G} networks \cite{shokri-ghadikolaei_millimeter_2015}. We foresee that for assistive health and beauty applications related to the proposed \gls{IoM} such as remote treatments and remote consultations, data transfer over the \gls{5G} network would be the most feasible since its main characteristics include a mobile access speed of 1Gbps, a fixed access speed of 1-10 Gbps and a data transmission delay of 1ms \cite{khanh_wireless_2022}. Another IoT ecosystem, the \gls{IoMusT} has been evaluated on a \gls{5G}-based system designed to support networked music performances (NMPs) in \cite{turchet_latency_2023}. The study suggests that while \gls{5G} networks hold promise for real-time interactive systems, like our foreseen \gls{IoM}, further improvements in terms of latency and reliability may be needed. Similarly, a future \gls{IoM} system could be evaluated on a private \gls{5G} deployment, focusing on latency and reliability of data packet exchanges over the \gls{5G} network. Such an evaluation could reveal how latency increases proportionally with the number of nodes and the presence of background traffic, and how reliability metrics may remain stable regardless of conditions. 
        
        The advent of the Tactile Internet, combining ultra-low latency with high availability, reliability, and security, opens up new opportunities for remote control of devices and could be a crucial enabler for advanced \gls{IoM} applications \cite{promwongsa_comprehensive_2021} \cite{simsek_5G-enabled_2016}. Given the importance of haptics and wearables in envisioned \gls{IoM} applications, the ultra-low latency of the Tactile Internet becomes crucial.
        The connectivity requirements of such systems are stringent. These arise from the necessity for immediate, reliable, and secure data transmission, especially when handling real-time, haptic-intensive tasks. The Tactile Internet, with its emphasis on ultra-low latency, can meet these demands. Thus, the advent of the Tactile Internet could not only enhance the capabilities of current \gls{IoM} applications but also pave the way for novel, haptic-intensive applications in the future.
        
        Furthermore, the convergence of \gls{IoT} and wireless connectivity is laying the groundwork for the development of "Smart Cities", where numerous connected devices can interact and exchange data \cite{rp_iot_2021}. Such an environment could provide a more seamless and integrated experience for users of an \gls{IoM} system, broadening the possibilities for assistive health and beauty applications.

    \subsection{Computational Architectures}

        Computational architectures and approaches are essential to the realisation of the \gls{IoM}, enabling seamless integration and efficient processing of data in real-time. In particular, edge computing, cloud computing, and distributed computing are integral to the functioning of \gls{IoM}, with each offering unique advantages and capabilities.

        \begin{itemize}
        
            \item \textbf{Edge computing} involves processing data near its source, reducing latency and bandwidth usage, and improving real-time interactions \cite{satyanarayanan_emergence_2017} \cite{shi_edge_2016}. In the context of \gls{IoM}, edge computing enables faster processing and analysis of user data collected by smart mirrors, providing real-time feedback and recommendations to users. For instance, when a user stands in front of a smart mirror, the data from the user's image can be processed immediately at the edge, enabling the mirror to provide immediate feedback or recommendations.

            \item \textbf{Cloud computing} provides scalable and flexible computing resources, data storage, and services over the internet \cite{bonomi_fog_2014}\cite{georgakopoulos_internet_2016}. In the \gls{IoM}, cloud computing can be used to store and process large amounts of user data collected by smart mirrors. For instance, cloud-based \gls{ml} algorithms can analyse this data to detect patterns and trends, informing personalised recommendations.

            \item \textbf{Distributed computing} involves dividing a larger computation problem into smaller parts that are solved concurrently \cite{escamilla-ambrosio_distributing_2018}. Distributed computing approaches can be used to manage the vast amounts of data generated by smart mirrors in the \gls{IoM}. By distributing the computation across multiple nodes, these systems can process large amounts of data more quickly and efficiently.
            
        \end{itemize}
    
        One of the challenges for the \gls{IoM} is to effectively integrate these computing approaches. The integration of edge, \gls{IoT}, and cloud computing in a distributed computing environment can provide a comprehensive solution \cite{el-sayed_edge_2018}. The use of edge computing allows for real-time data processing and interaction, while cloud computing offers scalable resources and services. Distributed computing ensures efficient processing of large data sets.
        
        Moreover, the integration of these computing architectures can support advanced features in the \gls{IoM}, such as real-time health monitoring, advanced image processing, and personalised recommendations. For example, edge computing can support real-time health monitoring by processing vital signs data immediately. Cloud computing can support advanced image processing and \gls{ml} algorithms by providing the necessary computing resources. Distributed computing can support personalised recommendations by enabling efficient processing of large user datasets \cite{mao_survey_2017} \cite{patel_contributor_nodate}.

    \begin{table*}[h]
        \centering
        \caption{Overview of Enabling Technologies and Their Potential for IoM Integration}
        \small
        \begin{tabular}{|p{2.5cm}|p{4cm}|p{4cm}|p{4cm}|}
        \hline
        \textbf{Category} & \textbf{Technologies} & \textbf{Applications} & \textbf{Potential for IoM Integration} \\
        \hline
        Skin Imaging Methodologies & Classification and feature extraction, segmentation, capacitive imaging & Disease detection, wearable skin sensors & Early diagnosis, treatment recommendations \\
        \hline
        Real-Time Health Monitoring & PPG, Infrared thermography, RF Sensing & Vitals tracking, disease monitoring & Continuous monitoring, remote healthcare \\
        \hline
        Computer Vision & Augmentation of cosmetics & Makeup applications, retail scenarios & Real-time makeup previews, personalised shopping \\
        \hline
        Personalised Recommendation Systems & Deep learning, user feedback & Product suggestions, routine planning & User-specific recommendations, adaptive learning \\
        \hline
        Wireless Connectivity in \gls{IoT} & 5G, LPWAN, WLAN, Tactile Internet & Wireless data transfer, remote control & Low latency, high data rates, scalability \\
        \hline
        Computational Architectures & Edge Computing, Cloud Computing, Distributed Computing & Data Processing, Scalable Resources & Real-time monitoring, Large dataset handling \\
        \hline
        \end{tabular}
        \label{tab:IoMtechintegration}
    \end{table*}

\section{IoM Scenarios and Use Cases}

    We expect that the realisation of the \gls{IoM} will allow many everyday activities and aspects of human life to be re-imagined with respect to a technological ecosystem of interactions between stakeholders within the IoM ecosystem. These stakeholders includes:
    
    \begin{itemize}
        \item Consumers: Individuals such as customers and patients who interact with the IoM system to receive personalised services and recommendations.
        \item Professionals: Specialists like dermatologists, makeup artists, and other professionals who provide expertise and services through the IoM platform.
        \item Service Providers: Entities such as businesses, retail stores, and healthcare providers that offer products and services integrated into the IoM system.
    \end{itemize}

\begin{figure*}[ht]
        \centering
        \includegraphics[width=\textwidth]{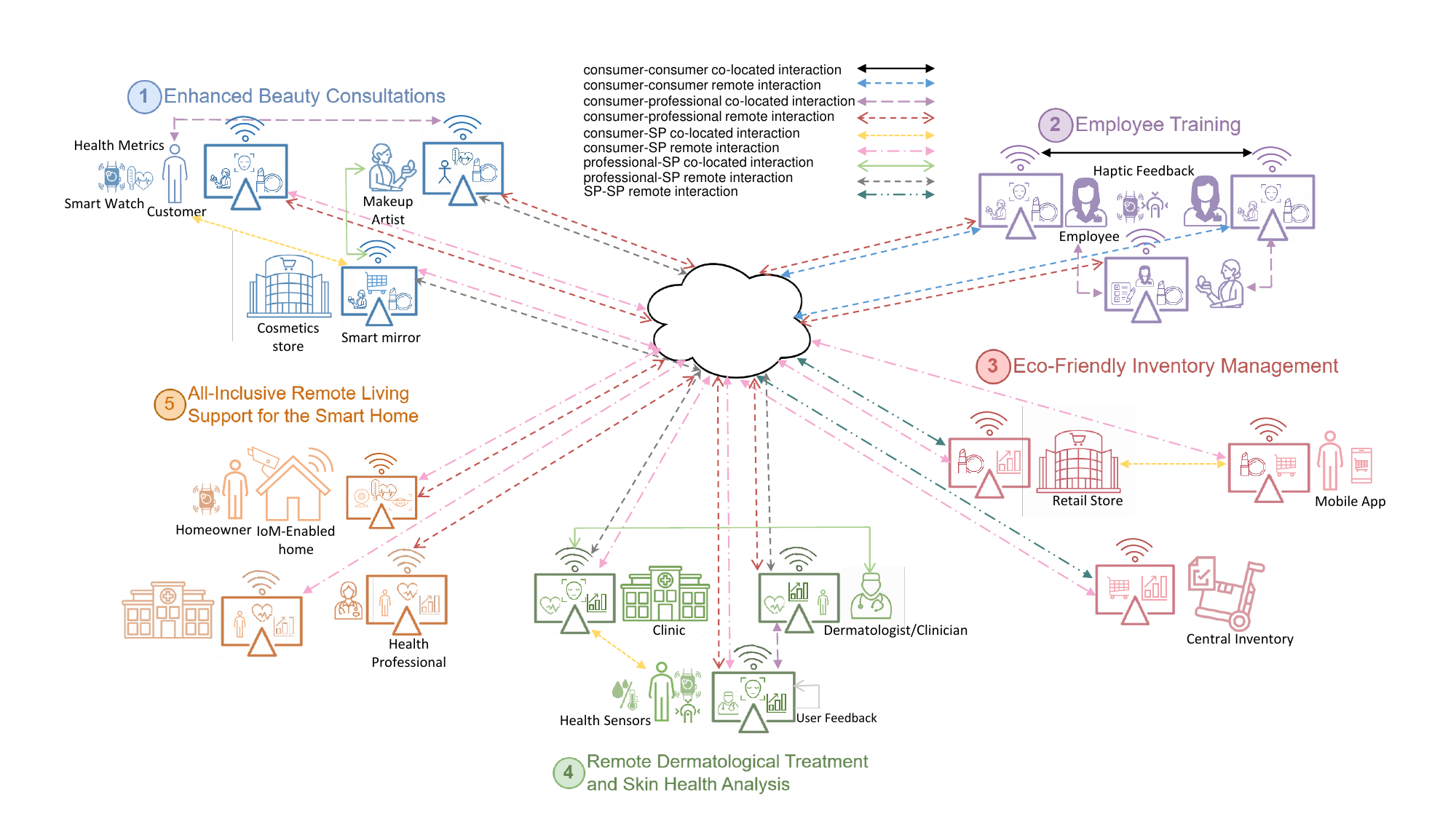}
        \vspace*{-0.7cm}
        \caption{Visualisation of IoM Use Cases and Possible Stakeholder Interactions}
        \label{fig:IoMusecase}
    \end{figure*}

    To present an idea and visualisation of what these interactions could be like and what the implications would be, five hypothetical use case scenarios that could potentially arise from the realisation of the \gls{IoM} are presented below as well as the corresponding missing capabilities and considerations. Following on from Fig \ref{fig:IoMinteractions}, a visualisation of these use case scenarios with all the relevant stakeholders and the possible interactions between them all within the ecosystem of the IoM are presented in Fig \ref{fig:IoMusecase}

    \subsection{\textit{Scenario 1 (Enhanced Beauty Consultations):}}
    Users entering beauty retail stores equipped with \gls{IoM} devices can approach and engage with interactive smart mirrors. This mirror, fitted with advanced image processing and computer vision capabilities, performs an immediate analysis of the customer's facial features, skin type, and complexion. Simultaneously, the smart mirror connects to the customer's wearable device, such as a smartwatch, to gather additional data such as skin moisture levels, heart rate, and body temperature. This data allows the mirror to provide enhanced personalised recommendations, taking into account the customer's physiological state. Once the initial analyses are completed, the customer can select various makeup products they wish to virtually try on, such as lipstick, eyeliner, or foundation. The smart mirror in real-time then augments the chosen products on the customer's face virtually, enabling them to see how each product complements their skin without physically applying it.
    The mirror can then also suggest products based on the customer's skin analysis, wearable data, and their selected preferences. It can propose products that are compatible with the customer's skin type, facial features, physiological state, or the look they aim to create. If the customer is satisfied with the virtual try-on, they can add the products to their shopping cart directly via the smart mirror interface. The results from the try-on are stored in the customer's user profile, allowing them to access their personalised recommendations and try-on results from any connected smart mirror in the \gls{IoM} ecosystem. This feature also ensures that their future suggestions are more personalised and accurate, as the system learns from their preferences over time. This scenario extends to remote situations as well where a user could interact with a beauty consultant remotely through the smart mirror. The consultant, using the customer's facial analysis, wearable data, and virtual try-on results, can offer personalised advice and product recommendations. 
    
    \subsection{\textit{Scenario 2 (Employee Training):}}
    Consider a beauty retail store equipped with \gls{IoM} devices where a new employee such as a Makeup Artist (MUA) is beginning their training. The trainee can engage with the smart mirror, which has been set up for a training session. The training session commences with the smart mirror displaying a live video feed of a senior artist demonstrating a makeup technique. As the senior artist explains and performs each step, the smart mirror overlays \gls{AR} guides on the live video feed. These guides assist the trainee by visually emphasising the crucial parts of the technique, such as the amount of product to use or the angle and direction of brush strokes. As the trainee replicates the demonstrated technique, the smart mirror uses its image processing and computer vision capabilities to analyse their actions in real time. If the trainee's application aligns with the demonstrated technique, they receive positive feedback. However, if the trainee deviates from the technique, the mirror provides corrective feedback in real time, and the trainee's  smartwatch can give haptic feedback as an immediate corrective signal. This immediate feedback helps the trainee to learn and adapt their techniques on the spot, enhancing learning efficiency. This concept extends to remote situations as well. Imagine a centralised training session where all the employees of a beauty brand across the country, or even globally, can participate. The smart mirror displays a live video feed of the brand's most senior artist demonstrating various makeup techniques. The employees can follow along on their local \gls{IoM} devices, receiving personalised feedback and haptic corrections as they practice each technique.
    After the training, the smart mirror can administer tests based on the techniques covered during the session. It evaluates the trainee's performance using the same real-time analysis capabilities, grading them based on their precision and adherence to the demonstrated techniques.

    \subsection{\textit{Scenario 3 (Eco-Friendly Inventory Management):}}
    In a beauty retail store equipped with \gls{IoM} devices where customers can interact with smart mirrors, every action taken is logged and analysed. This includes the products they choose to virtually try-on, the products they add to their shopping cart, the recommendations they accept or reject, and even the products they opt not to interact with. These interactions provide extensive data about customer preferences and the performance of different products. Simultaneously, the smart mirrors are also equipped with sensors that collect data over time. If a customer purchases a product and uses it regularly, the smart mirror can track changes in the customer's skin over time, providing valuable feedback on the product's effectiveness and long-term results. All this data can be aggregated and analysed to identify trends, preferences, and areas for improvement. Businesses can use these insights to inform their inventory management, ensuring popular products are adequately stocked to meet demand and reducing the chances of products being sold out. Moreover, by predicting product demand more accurately, businesses can reduce overproduction and wastage, contributing to more sustainable and eco-friendly practices. They can also optimise their supply chain, ensuring products are restocked efficiently, further reducing waste and inefficiencies. In remote settings, customers using smart mirrors at home can have their interactions and usage patterns logged and analysed. This data can offer businesses a broader understanding of how customers use their products in their daily routines, providing valuable insights for inventory planning and product development.


    \subsection{\textit{Scenario 4 (Remote Dermatological Treatment and Skin Health Analysis):}}
    Users with their skin health tracking wearable and smartwatch at home can be scanned by their smart mirror to analyse a range of skin parameters such as skin type, moisture levels, pore size, blemishes, wrinkles, and signs of ageing. Concurrently, the wearable device and smartwatch measure physiological data such as heart rate, body temperature, and hydration levels, providing a comprehensive context to the skin analysis. This combined data from the smart mirror, wearable device, and smartwatch is then transmitted remotely to a dermatologist for consultation. The dermatologist can view the detailed skin and physiological analysis on their end, enabling them to make an informed diagnosis and suggest personalised treatment plans, ranging from skincare product recommendations to lifestyle changes and medical treatments, based on the user's skin condition and overall health. The smart mirror can also offer \gls{AR} tutorials to the user. For instance, if the dermatologist prescribes a particular skincare routine, the mirror can display a step-by-step \gls{AR} guide illustrating how to correctly apply the recommended products. The mirror's AI capabilities can monitor the user's application, providing real-time feedback to ensure the correct treatment application. If the user applies a product incorrectly, the smartwatch can provide haptic feedback as an immediate corrective signal. Post consultation, the smart mirror and wearable devices continue to monitor the user's skin health and physiological data. This allows for tracking the effectiveness of the prescribed treatments, offering valuable feedback to both the user and the dermatologist. If the mirror or any wearable device detects significant changes in the skin's condition or the user's health, they can alert the user and the dermatologist through notifications and haptic feedback, enabling prompt adjustments to the treatment plan. Through this user feedback, the smart mirrors in the ecosystem can also learn from each other about different skin types and the effects of different treatments and products to provide improved recommendations ensuring diversity and inclusion.  
   
    \subsection{\textit{Scenario 5 (All-Inclusive Remote Living Support for the Smart Home):}}
    In a smart home equipped with \gls{IoM} devices, the smart mirror serves as a central control and monitoring hub and is integrated with various other smart devices in the home, such as security cameras, smart locks, thermostats, and health monitoring devices. The smart mirror is equipped with advanced AI capabilities that enable it to monitor and control various aspects of the home. For instance, it can display live feeds from security cameras, alert homeowners of any detected anomalies, and control smart locks, enhancing the home's overall security. The smart mirror can also detect anomalies within the home, such as a sudden temperature drop or an unusual increase in noise levels. These anomalies can trigger alerts to the homeowner, allowing them to take necessary actions promptly. Integrated with the home's smart devices, the smart mirror can also control various aspects of the home environment. For instance, it can adjust the thermostat based on the homeowner's preferences or turn off lights in unoccupied rooms, contributing to energy efficiency. In terms of health monitoring, the smart mirror can connect to wearable devices to track various health parameters such as heart rate, body temperature, and sleep patterns. This data can be displayed on the mirror for easy viewing and tracking over time. If the mirror detects any significant changes or concerning trends in these health parameters, it can alert the homeowner or even a healthcare provider. Furthermore, the smart mirror can serve as a personal assistant, providing reminders for appointments, displaying news updates, or even providing makeup and skincare recommendations based on the user's preferences and skin condition.

    \subsection*{\textbf{Unique Advantages}}
    While the scenarios presented above may share similarities with existing technologies, the IoM offers unique advantages that set it apart. The IoM is not designed to replace smartphones, tablets, or other smart devices, but rather to complement them by providing a specialised, visual interface that seamlessly integrates into daily life, particularly in health and beauty contexts to offer comprehensive monitoring, real-time full-body analysis, and personalised recommendations through an always-on display that enhances user experience beyond what handheld devices can provide. 

    The key differentiators of the IoM include its passive integration into daily routines, requiring minimal active engagement from users, and its role as a unified platform for health and beauty services, ensuring data consistency and interoperability. Unlike fragmented apps and services, the IoM provides a centralised ecosystem with a large, always-on visual interface that offers full-body visualisation and interaction, which is particularly suited for health and beauty applications. This design allows for hands-free operation, making it ideal for scenarios where manual interaction with devices is impractical.
    
    Furthermore, the IoM can facilitate more standardised and seamless integration of professional services, from beauty consultations to healthcare diagnostics. Its fixed position allows for consistent, long-term monitoring of user health and appearance, enabling more accurate tracking of changes over time. These features position the IoM as a novel, unified platform that addresses specific needs in the health and beauty domains, offering capabilities that go beyond what current fragmented solutions can provide. Table \ref{tab:iom-unique-advantages} provides a more comprehensive comparison between the IoM and existing systems and outlines the specific advantages of the IoM and how they address limitations in current devices.
    \begin{table}[ht]
    \centering
    \caption{Detailed Breakdown of IoM's Unique Advantages}
    \label{tab:iom-unique-advantages}
    \small
    \begin{tabular}{|p{0.2\textwidth}|p{0.4\textwidth}|p{0.3\textwidth}|}
    \hline
    \textbf{Feature} & \textbf{Description} & \textbf{Advantage} \\
    \hline
    Full-body, hands-free analysis & Comprehensive full-body scans and virtual try-ons without handheld devices. & Overcomes limitations of smaller, handheld devices for full-body assessments. \\
    \hline
    Passive, continuous monitoring & Monitors health and beauty metrics over extended periods without active user intervention. & Complements wearables, offering continuous monitoring without charging or wearing devices. \\
    \hline
    Multi-user, shared experience & Enables multi-user interactions and community-based features in shared spaces. & Facilitates shared experiences not easily achieved with personal devices. \\
    \hline
    Professional-grade imaging & Consistent lighting and high-resolution cameras for detailed skin analysis. & Surpasses varying quality of phone cameras and lighting conditions. \\
    \hline
    Contextual recommendations & Provides environment-aware health and beauty recommendations. & Offers more consistent environmental tracking than mobile devices. \\
    \hline
    Seamless professional integration & Standardised platform for remote professional consultations. & Provides more consistent, high-quality visual data than varied personal devices. \\
    \hline
    \end{tabular}
    \end{table}

    These unique capabilities position the IoM as a complementary technology to existing devices, filling crucial gaps in current health and beauty tech ecosystems. By working alongside and enhancing the capabilities of smartphones, tablets, and wearables, the IoM provides a more comprehensive and integrated approach to personal health and beauty management.

    \subsection*{\textbf{Missing Capabilities}}

    \begin{itemize}
        \item Scenario 1 illustrates a typical assistive beauty application of the \gls{IoM} but it relies heavily on advanced AI, AR, image processing, and computer vision technologies. As of now, the accuracy of these technologies may not be sufficient, and there could be discrepancies between the virtual try-on and the actual result. Moreover, implementing a personalised user profile system that provides accurate suggestions over time requires sophisticated \gls{ml} algorithms and substantial amounts of data, which could pose privacy and security concerns.

        \item Scenario 2 relies on seamless real-time integration between the trainee's movements and any corrective mechanisms. Another main challenge of this scenario is accurately detecting and evaluating the trainee's makeup techniques using computer vision. The variation in makeup application techniques can be subtle and complex, making it difficult for the system to correctly assess the trainee's performance. Moreover, the system must be trained with a large amount of data to accurately provide haptic feedback. This might be challenging due to the complex nature of makeup application techniques and the need for expert labelling. This scenario can also be extended to employee or student training for other areas such as dermatology or dentistry.

        \item Scenario 3 assumes that users will consistently interact with the smart mirror and that the data logged will accurately represent their preferences. However, user behaviour can be complex and unpredictable, and the metrics collected may not always reflect actual preferences or purchasing behaviour. Additionally, while the \gls{IoM} can help reduce waste by predicting demand more accurately, it cannot control external factors such as supply chain disruptions that can lead to product shortages or excess. The accuracy of such predictions will rely heavily on the realisation of smart cities.

        \item Scenario 4 holds promising potential for remote dermatological treatment by supporting the diagnostic process, treatment planning and treatment tracking. This scenario could also be extended to other aesthetic health domains such as dentistry and optometry. However, it may not be able to replace in-person consultations entirely because the smart mirror's analysis might not detect certain skin conditions that can be observed through a physical examination. Moreover, data privacy and security are significant concerns, especially when dealing with sensitive health data.

        \item Scenario 5 can enhance home security and convenience but it also introduces potential risks. The integration of multiple smart devices increases the attack surface for potential cyber threats, raising concerns about data security and privacy. Furthermore, the system's reliability is a significant factor - a system malfunction or false alarm could cause unnecessary distress or inconvenience for the homeowner. Ensuring the system's robustness, security, and reliability is essential for this scenario to be effective.
        
    \end{itemize}

\section{Current Challenges and Future Research Directions}

This section addresses the challenges associated with the development and realisation of the \gls{IoM}. It highlights key obstacles including communication protocol performance, system interoperability, multimodal content representation and data privacy and security. Possible solutions to these challenges that can support the practical realisation of \gls{IoM} systems have also been presented.
.

    \subsection{Low Latency and High Reliability}

    \begin{itemize}
        \item \textbf{Challenge}
        
        Real-time interactions would be fundamental to foreseen \gls{IoM} applications. Whether it is streaming a live makeup tutorial with \gls{AR} overlays, real-time health monitoring, or providing instant user feedback, low latency and high reliability are essential. The requirements for these systems bear a close resemblance to those for real-time gaming or video conferencing, with the need for high-quality, real-time data streams over both wired and wireless networks \cite{ma_iqbal_internet_2021}. 

        Achieving low latency and high reliability is a significant engineering challenge due to the strict requirements of network delay and transmission reliability. These requirements are set by the expectation of a high-quality interaction experience. A stable message reception rate and synchronisation between devices are crucial to maintaining a high level of user interaction \cite{chettri_comprehensive_2020}. For envisioned \gls{IoM} applications, the latency must be on the order of milliseconds. 
    
        Another challenge in communication services is the synchronisation of data streams produced by devices that do not share the same clock. Even if devices from different networks initially share the same clock, they require a re-synchronisation procedure from time to time. Several protocols have been proposed to achieve such synchronisation, but existing methods are insufficient as they do not meet ultra low latency requirements \cite{schwartz_modern_2021}.

        \item\textbf{Future Research Direction}
        
        To address this challenge 'physical layer synchronisation' is an emerging approach at the wireless communication interface, where the waveform is designed to carry both data and synchronisation information simultaneously \cite{ji_ultra-reliable_2018}. Another approach being explored is the optimisation of existing protocols which involves adapting the communication protocol parameters that are free to optimise the transmission of data. This has been used to support various protocols over Bluetooth or to optimise the protocol parameters of Wi-Fi for various applications \cite{elhabyan_coverage_2019}. However, these attempts are still not satisfactory, mostly due to the inherent limitations of the physical layer of the communication protocol, necessitating further research and development \cite{ji_ultra-reliable_2018}. 
    
        The envisioned Tactile Internet is expected to solve these issues by providing communication networks, both wireless and wired, capable of ensuring ultra-low latency communications, with end-to-end delays on the order of a few milliseconds \cite{maier_tactile_2016}. Along the same lines, the use of edge computing technologies is expected to play a relevant role in reducing latency and bandwidth pressure by offloading computation from the cloud to the edge of the network \cite{escamilla-ambrosio_distributing_2018}. 
        
        The primary data types of the IoM are expected to be image and video meaning that the demands on bandwidth and latency are even more stringent. The transmission of high-resolution images and video streams requires significant bandwidth and low latency to ensure a seamless user experience. This can be ensured through the development of advanced image and video compression techniques, as well as efficient coding and transmission protocols.
        
        Interference management is also an important consideration for low latency, specifically ultra-reliable low-latency communications (URLLC) which is a key deliverable of 5G networks and beyond 5G networks. Effective interference management is necessary for meeting the requirements of URLLC and ensuring a seamless user experience \cite{siddiqui_urllc_2023}. Many new solutions and insights are being explored in this area and an integrated research direction focusing on interconnected IoMs will support the realisation of a dense IoM ecosystem.
        
    \end{itemize}

    \subsection{Interoperability and Standardisation} 
    
    \begin{itemize}
        \item \textbf{Challenge}
    
        Interoperability and standardisation are fundamental to the successful implementation of the \gls{IoM}. These factors ensure interoperability, compatibility, reliability, and effective operations on both local and global scales. Despite their significance, much work remains to be done to realise fully standardised formats, protocols, and interfaces in the overarching \gls{IoT} field to provide more inter-operable systems \cite{ma_iqbal_internet_2021}.

        In the conceptualised \gls{IoM} ecosystem, various types of devices are expected to be utilised for different purposes, both in co-located and remote scenarios. These devices will be used to generate, track, and interpret multimodal content, primarily comprised of images and videos. They will need to dynamically discover and spontaneously interact with heterogeneous computing resources, physical resources, and digital data. This interconnection poses specific challenges, including the need for ad-hoc protocols and interchange formats for image and video data relevant to the \gls{IoM}, specifically how these visual data types are encoded, stored, and shared so that different devices and systems in the \gls{IoM} can understand and display them correctly. Image and video data in the \gls{IoM} could be transmitted over a WLAN using standard protocols such as Wi-Fi and Bluetooth, which are commonly provided by a variety of smart devices. However, the challenge lies in establishing a standard that can facilitate interoperability across heterogeneous devices without sacrificing flexibility as well as accounting for synchronisation aspects.

        \item\textbf{Future Research Direction}
        
        Semantic web technologies are a promising research avenue for addressing interoperability issues \cite{hitzler_foundations_2009}. This would ensure that data is not only interoperable on a syntactic level but also on a semantic level, meaning that the significance and context of the data are preserved and understood across different parts of our ecosystem. This would involve developing appropriate ontologies to describe the processes mediated by \gls{IoM} devices with a focus on real-time applications since these techniques require large segments of data to make predictions and this can be a challenge in a real-time context. 
        
        To address standardisation challenges, future research may focus on the adaptation of existing multimedia standards, such as those from MPEG, which provide a solid foundation for managing complex multimedia data. Such a format would accommodate the multi-modal nature of the \gls{IoM}, which requires standards for formats that account not only for the transmission of images and videos but also for associated metadata. In the context of assistive health and assistive beauty applications, standardisation plays a pivotal role. Devices used in these fields, such as health monitors and \gls{AR} mirrors, need to interact seamlessly with each other and with other systems. Therefore, the establishment of standard protocols and interchange formats is crucial. These would enable these devices to share and interpret data, leading to enhanced user experiences and more effective operations. Moreover, the nature of health and beauty applications requires strict data privacy and security standards. Data transmitted in these applications is often sensitive, and its misuse could lead to severe repercussions. Therefore, in addition to operational standards, rigorous data privacy and security standards will also need to be established.

        In addition, leveraging industry standards and best practices can also enhance the development and adoption of the IoM framework. For example, the TM Forum \cite{TM}, a global industry association which provides a comprehensive suite of frameworks and best practices that can be particularly valuable in this context. The TM Forum's frameworks, such as the Business Process Framework (eTOM) and the Information Framework (SID), offer structured methodologies for managing business processes and information systems in a scalable and efficient manner. Integrating these frameworks into the IoM concept can ensure better alignment with industry standards, enhance interoperability, and provide a robust foundation for the future development of the IoM ecosystem. Exploring and adopting relevant aspects of the TM Forum's frameworks can thus contribute to overcoming some of the challenges identified and support the realisation of a more resilient and standardised IoM infrastructure.
        
    \end{itemize}
        
    \subsection{Representation and Analysis of Multimodal Content}

    \begin{itemize}
        \item \textbf{Challenge}
        
        Different types of signals are expected to be exchanged within the \gls{IoM} including those associated with various sensory modalities, such as image data, \gls{rf} signals, environmental data, temperature data, and data from wearables. They will be dynamically captured through various sensors embedded in \gls{IoM} devices, providing varied information about the environment and user behaviour. For example, image-related information could reveal details about user interactions with \gls{IoM} devices or their environment. \gls{rf} signals could provide information about device connectivity and communication. Environmental data could provide insight into the conditions in which \gls{IoM} devices operate, and wearable data can provide a wealth of information about the user's physiological responses and activities. This multimodal data can be harnessed to study user behaviour, environmental changes, and the effects of these factors on the functioning and efficiency of \gls{IoM} devices. However, the major challenge the IoM will encounter with varied data types is the development of systems capable of interpreting multimodal information. 

        \item\textbf{Future Research Direction}
        
        The research direction taken to address the above issue should include the representation, translation and fusion of multimodal data exchanged in the envisioned \gls{IoM} \cite{holthaus_how_2016}. Where representation involves determining how to represent and summarise the data, translation or mapping involves determining how to translate data from one modality to another and fusion involves joining information from two or more modalities to perform a prediction. A possible research avenue includes the integration of massive machine-type communications (mMTC) \cite{dutkiewicz_massive_2017} in the IoM ecosystem to enable efficient, reliable, and secure exchange of health data wearables and medical data. This would facilitate not only the real-time transmission of high-resolution skin images and associated metadata but also the integration of diverse data streams. 
    \end{itemize}
    
    \subsection{Privacy and Security Issues}
    
    \textit{ii) Security Issues}
    
    As with other \gls{IoT} ecosystems, we predict that the \gls{IoM} ecosystem will face substantial challenges related to privacy and security issues. These challenges are more significant given the diverse range of data types collected and processed within the \gls{IoM} such as personal image data, \gls{rf} signals, environmental data, and wearable device data. Ensuring the secure and ethical operation of the \gls{IoM} requires careful attention to these issues.
    
    Devices within the \gls{IoM} ecosystem will need to communicate wirelessly and are therefore exposed to the associated security risks. More specifically, in the transition to next-generation networks like 6G, the interconnected nature of the\gls{IoM} will require embedded security automation and thorough assurance processes to enforce security policies effectively. Trustworthy systems have been defined as a key paradigm of the emerging 6G \cite{mohjazi_journey_2023}, highlighting the importance of secure and resilient network platforms particularly when dealing with complex environments. The seamless incorporation of AI and ML into network systems raises critical issues of transparency and trust. Given the complexity and non-transparent nature of these technologies, validating their security is currently limited to output testing against known inputs. To ensure that AI and ML can be safely integrated into essential services and infrastructure, there is a need to establish robust methods for assessing AI security. This involves confirming the confidentiality of data used in AI processes, preventing the exposure of sensitive information, verifying the dependability and clarity of AI decision-making, and safeguarding the overall integrity and resilience of the systems.

    A potential direction to ensure security is robust encryption which can be achieved through various cryptographic techniques. Implementing such robust encryption presents two primary challenges. Firstly for \gls{IoM} devices to be powerful enough to support encryption. and secondly, the use of more efficient, less energy-consuming cryptographic algorithms \cite{rajesh_chapter_2021} to be deployed. Additionally, blockchain technology can be used to create a transparent and immutable record of transactions and interactions to add another layer of security within the \gls{IoM} ecosystem such as being currently developed within the \gls{IoMusT} ecosystem \cite{turchet_blockchain-based_2022}. A uniform security standard will also need to be established across the \gls{IoM} framework and industry to ensure the safety of the data collected by \gls{IoM} devices.

    \textit{ii) Privacy Issues}
    
    Given the personalised nature of the \gls{IoM}, there will also be a need to implement transparent privacy mechanisms across a diverse range of \gls{IoM} devices and supporting platforms. Similar to protection from security risks, privacy in the \gls{IoM} could be preserved by leveraging encryption and blockchain technologies, where encryption will protect the data while in transit, and blockchain will ensure data integrity.
    
    Issues of data ownership must also be addressed to ensure the privacy and data integrity potential \gls{IoM} users participating in \gls{IoM}-enabled activities. \gls{IoM} devices could be equipped with machine-readable privacy policies, allowing devices to check each other's privacy policies for compatibility before communicating. 

    To counteract any data security threats, privacy-preserving computational approaches can also be integrated into the IoM ecosystem. Since many of the services of the \gls{IoM} are enabled by machine learning, alternative approaches such as federated learning (FL) can be employed to ensure data privacy by locally training models where the data is generated \cite{al-quraan_edge-native_2023}. This technique also reduces the load on network resources as well as computational cost. Since the development of FL is in its early stages, its future research directions can be aligned and integrated with the IoM to support the large envisioned scale, network demands and integration of AI in our ecosystem.

\section{Conclusion}

This paper proposed the innovative framework of the \gls{IoM} which emerges as an ecosystem under the umbrella of \gls{IoT}. The \gls{IoM} envisions an interconnected network of smart mirrors, which not only reflects the user but also superimposes layers of digital information and personalised user metrics, enhancing daily routines for individuals across various demographics and settings. We have highlighted that beyond the user's reflection, \gls{IoM} smart mirrors can offer many unique services in a single platform such as health insights, employee training or even inventory management. These capabilities could profoundly influence personal care routines, fashion choices, health monitoring, and even mental well-being. The \gls{IoM} framework has been defined to be composed of three major elements, smart mirrors, connectivity and application and services. We have illustrated that the realisation of the \gls{IoM} and its elements will be enabled by the integration of many state-of-the-art technologies such as existing standalone smart mirror technologies, health monitoring and wireless connectivity. To support this concept and to highlight future possibilities, we also presented unique envisioned scenarios that could arise from the \gls{IoM} by including examples from assistive beauty, assistive health and assistive living. This was followed by a discussion of the probable challenges we foresee the \gls{IoM} ecosystem to face and potential solutions as well future research directions that can support and be aligned with the IoM vision. We anticipate that the potential of our conceptualised framework is vast by bringing a passive everyday object to the forefront of interactive technology.  The \gls{IoM} concept presents a novel unified platform for not only assistive beauty and health but also assisted technologies as a whole and potentially the future vision of smart cities. 

\bibliographystyle{elsarticle-num}
\bibliography{main}

\end{document}